\documentclass{article}
\usepackage{spconf, amsmath, graphicx}
\usepackage{subcaption}
\usepackage{soul}
\usepackage{cite}

\usepackage{enumitem}
\setlist{nosep, leftmargin=14pt}

\newlength{\bibitemsep}
\newlength{\bibparskip}
\let\oldthebibliography\thebibliography
\renewcommand\thebibliography[1]{
  \oldthebibliography{#1}
  \setlength{\parskip}{\bibitemsep}
  \setlength{\itemsep}{\bibparskip}
}

\usepackage{mwe}
\usepackage{amssymb}
\usepackage{bm}
\usepackage{xcolor}

\newcommand{\R}{\mathbb{R}}
\newcommand{\C}{\mathbb{C}}
\newcommand{\refeq}[1]{Eq.\,\eqref{#1}}
\newcommand{\reffig}[1]{Fig.\,\ref{#1}}

\title{DELTA-MRI: Direct deformation Estimation from LongiTudinally Acquired k-space data}

\name{J. Renders$^{\star  \ddagger}$, B. Shafieizargar$^{\star \ddagger}$, M. Verhoye$^{\dagger \ddagger}$, J. De Beenhouwer$^{\star  \ddagger}$, A.J. den Dekker$^{\star \ddagger}$, J. Sijbers$^{\star \ddagger}$}
\address{$^{\star}$imec-Vision Lab, Department of Physics, University of Antwerp, Antwerp, Belgium, \\
$^{\dagger}$Bio-Imaging Lab, Department of Biomedical Sciences, University of Antwerp, Antwerp, Belgium, \\
$^{\ddagger}$$\mu$NEURO Research Centre of Excellence, University of Antwerp, Antwerp, 
Belgium}

\begin{document}

\maketitle

This work has been submitted to the IEEE for possible publication. Copyright may be transferred without notice, after which this version may no longer be accessible.

\begin{abstract}
Longitudinal MRI is an important diagnostic imaging tool for evaluating the effects of treatment and monitoring disease progression. However, MRI, and particularly longitudinal MRI, is known to be time consuming. To accelerate imaging, compressed sensing (CS) theory has been applied to exploit sparsity, both on single image as  on image sequence level. State-of-the-art CS methods however, are generally focused on image reconstruction, and consider analysis (e.g., alignment, change detection) as a post-processing step.

In this study, we propose DELTA-MRI, a novel framework to estimate longitudinal image changes {\it directly} from a reference image and subsequently acquired, strongly sub-sampled MRI k-space data. In contrast to state-of-the-art longitudinal CS based imaging, our method avoids the conventional multi-step process of image reconstruction of subsequent images, image alignment, and deformation vector field computation. Instead, the set of follow-up images, along with motion and deformation vector fields that describe their relation to the reference image, are estimated in one go. Experiments show that DELTA-MRI performs significantly better than the state-of-the-art in terms of the normalized reconstruction error. 
\end{abstract}
\section{Introduction}
\label{sec:intro}

In longitudinal MRI studies, repeated MRI scans are performed in specific clinical scenarios, such as follow up of patients \cite{Angelini12},  therapy response assessment \cite{Weizman14}, and large-scale longitudinal studies  \cite{Schilling22}. They constitute one of the most efficient tools to track pathology changes and to evaluate treatment efficacy in diseases. In such studies, patients are scanned on a regular basis, which calls for efficient longitudinal scan protocols.

Conventional longitudinal MRI workflows consist of multiple steps. First, complex-valued image data is acquired (in k-space) at different time points, from which (also complex-valued) images are individually reconstructed. Next, the magnitudes of the resulting images are mutually geometrically aligned. Finally, local structural deformations are detected by computing deformation vector fields, allowing for voxel-wise deformation analysis \cite{Dufresne20}. 

Unfortunately, conventional longitudinal MRI workflows are characterized by a low time efficiency. To accelerate imaging, techniques based on compressed sensing (CS) theory have been applied that exploit sparsity in some spatial transform domain and reconstruct images from substantially less data (through k-space sub-sampling) \cite{Lustig07}. Furthermore, since changes between two subsequent images in a longitudinal MRI study are generally small (apart from global transformations), temporal redundancy can be additionally exploited to shorten image acquisition time.  Indeed, CS-based methods have successfully been applied to also accelerate longitudinal MRI scanning by extending spatial regularization with temporal regularization \cite{Weizman15, Ye19, asif2013motion, jung2009k}. Such methods, however, require careful tuning of multiple regularization parameters to balance sparsity constraints (both in spatial and time domain) and data consistency, leading to impractical methods. Alternatively, deep learning methods have been investigated to automatically detect and exploit the distribution of expected images in both the spatial \cite{Tezcan19} and temporal \cite{yoo2021time, bustin2020compressed} domain. However, most data driven methods strongly rely on extensive training with a large number of datasets, which is especially non-trivial to obtain for longitudinal MRI applications.

In this paper, we propose DELTA-MRI, a novel framework to estimate longitudinal image deformations {\it directly} from a reference image and subsequently acquired, strongly sub-sampled MRI k-space data. In contrast to state-of-the-art longitudinal MR image reconstruction methods, which rely on the image reconstruction followed by deformation vector field estimation, our method avoids this multi-step process. Instead, the set of follow-up images, along with motion parameters and deformation vector fields that describe their relation to the reference image, are estimated in one go. We will show that DELTA-MRI performs significantly better compared to state-of-the-art CS based longitudinal MRI methods in terms of the normalized reconstructed error. 

\section{methods}
Let $\bm{x}_1=\{x_{1,k}\}, \bm{x}_2=\{x_{2,k}\} \in \C^N$ be two complex valued 3D MRI images with $N$ voxels in the image domain, of which the elements can be expressed in polar form as
\begin{equation}
    x_{j,k} = r_{j,k} e^{i\phi_{j,k}} \quad,\ \textrm{with}\ j = 1, 2 \ ;\ k=1,...,N
\end{equation}
and with $\bm{r}_j=\{r_{j,k}\} \in \R^N$ and $\bm{\phi}_j=\{\phi_{j,k}\} \in \R^N$ the magnitude and phase of $\bm{x}_j$, respectively.
Furthermore, let $\bm{d}_1, \bm{d}_2 \in \C^N$ be the k-space (Fourier) representations of $\bm{x}_1, \bm{x}_2$, respectively: $\bm{d}_j = \bm F \bm{x}_j$, with 

$\bm F \in \C^{N \times N}$
the discrete Fourier transform operator. 

Finally, let  $\bm{v} \in \R^{N \times 3}$ be a deformation vector field (DVF), comprised of one 3D vector per voxel, describing the geometrical deformation from $\bm{r}_1$ to $\bm{r}_2$, and let $ W(\cdot, \bm{v})$ denote the image warping operator \cite{renders2021adjoint} along 
$\bm{v}$ that (approximately) yields $\bm{r}_2$ when applied to $\bm{r}_1$:
\begin{equation}
    \label{eq:real_equation}
     W(\bm{r_1}, \bm{v}) = \bm{r}_2 \quad .
\end{equation}
We propose the estimation of $\bm{v}$ directly from k-space measurements of $\bm{x}_1$ and $\bm{x}_2$. These measurements, which will be denoted by $\{\tilde{\bm d_j}\}_{j=1}^2 \in \C^{n_j}$, 
are noise disturbed and can be modelled as:
\begin{equation}
    \tilde{\bm d_j} =\bm{S}_j \bm{F} \bm{x}_j + \bm \varepsilon_j, \quad \textrm{with}\ j = 1, 2 \ 
\end{equation}
where $\bm \varepsilon_j \in \C^{n_j}$ is an additive noise contribution, modeled as a zero-mean, complex-valued Gaussian random variable, and $\bm{S}_{j}\in\{0,1\}^{n_j\times N }$, with $n_j \leq N$, is  a sub-sampling operator that selects the k-space points acquired for the $j$-th image. 
For fully sampled k-space data, denoted by $\bm {\underline{\tilde d}}_j$, the operator $\bm{S}_j$ corresponds with the identity matrix $\bm I \in \R^{N \times N}$. 

In what follows, it is assumed that sufficient k-space data $\tilde{\bm d_1}$ is acquired to allow a high quality reconstruction $\hat{\bm r}_1$ of the reference magnitude image $\bm r_1$. If fully sampled data $\bm {\underline{\tilde d}}_1$ is available, such an estimate can be calculated as 
$\bm {\hat{r}}_1=|\bm{F}^{-1} \underline{\tilde {\bm d}}
_1|$, with $|\cdot|$ the point-wise modulus operator.
Furthermore, it is assumed that the phase $\bm{\phi}_2$ is slowly varying and that the low frequencies in k-space are sufficiently sampled to obtain a good estimate of $\bm{\phi}_2$ by calculating $\hat{\bm \phi}_2 =\angle (\bm F^{-1} \tilde{\bm{d}}_2)$, where $\tilde{\bm{d}}_2$ is zero-filled at the non-sampled indices and $\angle(\cdot)$ denotes the phase operator.
Under these assumptions, $\bm{v}$ can be directly estimated
by minimizing a regularized least-squares functional
without requiring the reconstruction of $\bm{x}_2$:
\begin{equation}
\label{eqn:estimator_v}
   \hat{\bm{v}}=  \arg \min_{\bm{v}} \left( \lVert \tilde{\bm{d}_2}-\bm{S}_2 \bm{F} W(\bm{\hat{r}_1}, \bm{v})e^{i\hat{\bm{\phi}}_2} \rVert_2^2 + \lambda R(\bm{v})\right),
\end{equation}
with 
$\lambda R(\bm{v})$ a regularization term with weight $\lambda$ that
imposes prior knowledge about
$\bm{v}$. 
In this work, 
$R(\bm{v})$ is chosen equal to $\|\nabla \bm v\|_2$, with $\nabla$ the gradient operator. This choice
reflects the assumption that $\bm{r}_2$ only differs from $\bm{r}_1$ by a local, smooth deformation, for which this regularization term is small.

In practice, $\bm{x}_2$ will likely also differ from $\bm{x}_1$ by a rigid transformation, on top of the local deformation. This transformation can be included in the DVF $\bm{v}$, but then $\bm{v}$ will no longer be mostly zero. Alternatively, the rigid transformation can be isolated from the DVF, and represented by rotation and translation parameters $\bm{\theta}, \bm{t} \in \R^3$. 
\refeq{eqn:estimator_v}
can then be extended as follows:
\begin{eqnarray} \label{eqn:joint_estimator}
   \{\bm{\hat  v, \hat \theta, \hat t}\}&=&\arg \min_{\bm{\theta}, \bm{t}, \bm{v}} ( \lVert  \tilde{\bm{d}}_2 - \bm{SF} W(\hat{\bm{r}}_1, \bm{\theta}, \bm{t, \bm{v}})e^{i\hat{\bm{\phi}}_2} \rVert_2^2 \nonumber \\ 
    &&+ \lambda R(\bm{v})) \quad
    ,
\end{eqnarray}
where $W(\cdot, \bm{\theta}, \bm{t, \bm{v}})$ is an image warping operator that rotates the image according to $\bm{\theta}$, translates the image according to $\bm{t}$ and then applies the local deformation described by $\bm{v}$. The optimization problem of \refeq{eqn:joint_estimator} can be solved using the cyclic Block Coordinate Descent (cBCD) method \cite{Beck13}, where the first block consists of the parameters $\bm{\theta}$ and $\bm t$ and the second block consist of $\bm{v}$.  In what follows, the initial estimates of  $\bm{v}$, $\bm t$ and $\bm \theta$ will be denoted as $\bm{v}_0$, $\bm{t}_0$ and $\bm{\theta}_0$, respectively.

\section{Experiments}
\label{sec:experiments}
High-resolution, fully-sampled k-space data was acquired using a Cartesian sampling scheme of a male C57BL/6 wild-type mouse using a 9.4T Biospec 94/20 USR horizontal MR system (Bruker Biospin MRI, Ettlingen, Germany), with a mouse head 2$\times$2 array cryo-coil (Bruker Biospin MRI, Ettlingen, Germany), using a 3D T2-weighted Turbo-RARE sequence with FOV: $20\times 15\times 10$ mm$^3$,
image acquisition matrix: $256\times 192\times 128$, 0.078 mm isotropic voxels, TE/TR: 67.2/1800 ms, RARE factor: 12. The fully-sampled multi-coil k-space data was then averaged across the coils to obtain
$\underline{\tilde{\bm{d}}}_2$. 

\begin{figure}[hbt]
    \centering
    \begin{subfigure}[t]{2.8cm}
    \includegraphics[width=2.8cm]{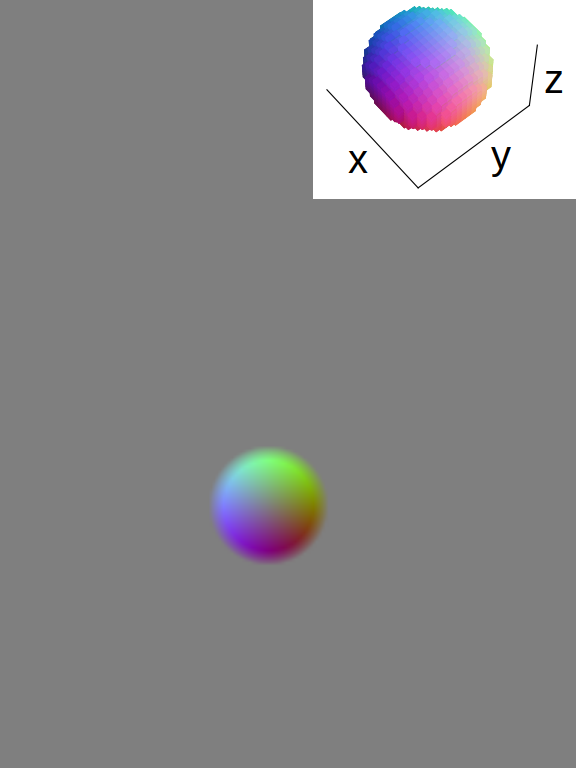}
    \caption{Ground truth DVF}
    \label{fig:GT_DVF_color}
    \end{subfigure}
    \begin{subfigure}[t]{2.8cm}
    \includegraphics[width=2.8cm]{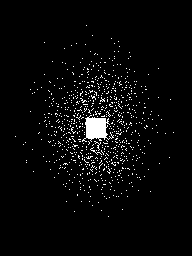}
    \caption{Gaussian 1\%}
    \label{fig:normal1}
    \end{subfigure}
    \begin{subfigure}[t]{2.8cm}
    \includegraphics[width=2.8cm]{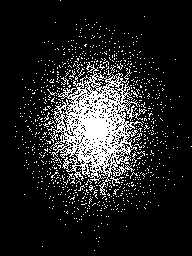}
    \caption{Gaussian 5\%}
    \label{fig:normal2}
    \end{subfigure}
    \caption{Ground truth DVF (a) Sampling schemes (b) and (c).}
    \label{fig:sampling_schemes} 
\end{figure}

\begin{figure}[htb]
    \centering
    \begin{subfigure}[t]{2.8cm}
    \includegraphics[width=2.8cm]{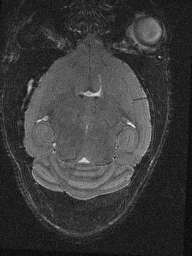}
    \caption{$\bm{r}_1$}
    \label{fig:r1}
    \end{subfigure}
    \begin{subfigure}[t]{2.8cm}
    \includegraphics[width=2.8cm]{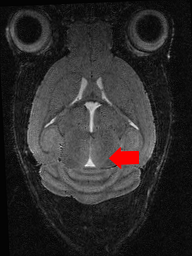}
    \caption{$\bm{r}_1$ aligned}
    \label{fig:r1aligned}
    \end{subfigure}
    \begin{subfigure}[t]{2.8cm}
    \includegraphics[width=2.8cm]{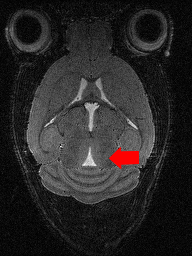}
    \caption{$\bm{r}_2$}
    \label{fig:r2}
    \end{subfigure}
    \caption{Reconstructions from fully sampled k-space data.}
    \label{fig:r1_and_r2} 
\end{figure}

Based on the experimentally acquired dataset, simulation experiments were setup to validate DELTA-MRI. First, a complex valued  image ${\bm x}_2=\bm{F}^{-1} \underline{\tilde{\bm{d}}}_2$ was reconstructed, with corresponding magnitude ${\bm r}_2$ and phase $\bm{\phi}_2$. 
From $\bm{r}_2$, a predefined rigid transform ($\bm{\theta}$, $\bm t$), and a predefined DVF $\bm{v}$, a magnitude image $\bm{r}_1$ was constructed 
using an inverse image warp, such that
$W(\bm r_1, \bm \theta, \bm t, \bm v) = \bm r_2$.
 \reffig{fig:GT_DVF_color} shows a cross section of the ground truth DVF $\bm v$.   The ground truth parameter 
  values of $\bm{\theta}$ and $\bm t$ were given by $(2.9^\circ, 4.0^\circ, 5.7^\circ)$ and $(-6, -5, -4.5)$ pixels, respectively. 
\reffig{fig:r1}, \ref{fig:r1aligned}, and \ref{fig:r2} show cross sections of $\bm{r}_1$, its fully aligned version $W(\bm {r}_1, \bm \theta, \bm t, \bm 0)$, and $\bm{r}_2$, respectively. Since DELTA-MRI does not depend on the phase of the reference image, its evaluation does not require the simulation of $\bm{\phi}_1$. However, as will be described below, DELTA-MRI was benchmarked against a state-of-the-art method that does depend on $\bm{\phi}_1$ and assumes it to be similar to $\bm{\phi}_2$.
Therefore, a complex valued image ${\bm x}_1$ with magnitude $\bm{r}_1$ was constructed, by setting $\bm {\phi}_1 = \bm{\phi}_2$. 
This complex image was polluted with complex, Gaussian distributed white noise with standard deviation equal to 4\% of the average foreground value of $\bm{r}_1$. We denote this noise disturbed image by $\hat{\bm {x}}_1$ and its magnitude and phase by $\hat{\bm {r}}_1$ and $\hat{\bm{\phi}}_1$, respectively. 
Next, a sub-sampling operator $\bm S_2$ was applied that under-samples $\underline{\tilde{\bm {d}}}_2$ by selecting a given percentage of the k-space data points. In particular, a small cubic central area was complemented by k-space points that were randomly drawn from a central Gaussian distribution with a standard deviation that corresponds with a quarter of the maximum frequency. 
Illustrative examples of the thus designed sampling schemes are shown in \reffig{fig:normal1} and \reffig{fig:normal2}.
Finally,  $\bm v$, $\bm{\theta}$ and $\bm t$, were estimated from the under-sampled k-space data $\tilde{\bm {d}}_2$ using the estimator described by \refeq{eqn:joint_estimator}. The 
experiments were repeated for different sub-sampling percentages. 

\textbf{Implementation details:}
The optimization task of \refeq{eqn:joint_estimator} was carried out using a cBCD approach with 
$\bm{v}_0$, $\bm{t}_0$ and $\bm{\theta}_0$ all 
zero. The value of $\lambda$ was set to $\lambda_0 \cdot 5 \cdot 10^{-5}$, where $\lambda_0 = \lVert \tilde{\bm{d}_2} \rVert_2^2$ maintains the regularization strength when varying the sampling scheme. This value was obtained by empirically minimizing the reconstruction error at 1\% Gaussian sub-sampling. In the first block, $\bm{\theta}$ and $\bm t$ were optimized with the \mbox{L-BFGS-B} method of SciPy 1.9.3 \cite{2020SciPy-NMeth}, with default stopping criteria and bounds $\bm t \in [-20, 20]^3, \bm{\theta} \in [-0.3 \text{ rad}, 0.3 \text{ rad}]^3$. In the second block,  $\bm{v}$ was optimized with the Barzilai-Borwein gradient method \cite{barzilai1988two} with 2000 iterations to ensure convergence. The derivatives of the objective function towards the deformation parameters were computed using ImWIP \cite{renders2022imwip, renders2021adjoint}. Because the deformation is local, the initial estimate
$\bm{v}_0$ was close enough to give an accurate estimation of $\bm{\theta}$ and $\bm t$. Therefore, each block was optimized only once. It was verified that adding more cycles did not change the results significantly.

\textbf{Benchmarking:} DELTA-MRI was benchmarked against
the Temporal Compressed Sensing MRI  (TCS-MRI) method \cite{Weizman15},
which regularizes the reconstruction of the complex image $\bm{x}_2$ in the temporal and wavelet domain, by solving
\begin{eqnarray}
\label{eqn:TCS-MRI}
\bm{\hat x}_2 &=&  \arg \min_{\bm{x}_2} ( \lVert \tilde{\bm{d}_2}-\bm{S}_2 \bm{F} \bm{x}_2 \rVert_2^2   \nonumber \\
&& + \lambda_1 \lVert \text{diag}(\bm{w}) (\hat{\bm{x}}_1 - \bm{x}_2) \rVert_1 + \lambda_2 \lVert \bm{\Psi} \bm{x}_2 \rVert_1 ),
\end{eqnarray}
with $\bm{\Psi}$ the Daubechies-4 wavelet transform and $\bm{w}$ a weight vector that controls the demand for similarity between $\bm{x}_1$ and $\bm{x}_2$, enforcing sparsity only in regions where $\bm{x}_1$ and $\bm{x}_2$ are similar. In our experiments, $\bm{w}$ was fixed to the support of the ground truth DVF. 
 
It should be noted that TCS-MRI assumes that the rigid transformation aligning $\bm{x}_1$ and $\bm{x}_2$ is known.
DELTA-MRI
does not make this assumption, and instead estimates the rigid transformation, along with $\bm v$,
from the available, sub-sampled data. In each experiment, the rigid transformation estimated by DELTA-MRI was used to align the data before using TCS-MRI. Furthermore, the regularization weights $\lambda_1$ and $\lambda_2$ in  \refeq{eqn:TCS-MRI} were set to $\lambda_0 \cdot 5.85 \cdot 10^{-9}$ and $\lambda_0 \cdot 3.63 \cdot 10^{-10}$, respectively, where $\lambda_0 = \lVert \tilde{\bm{d}_2} \rVert_2^2$ maintains the regularization strength when varying the sampling scheme. These values were obtained by empirical tuning to optimize reconstruction quality at 1\% Gaussian sampling.

To evaluate how much information on $\bm{r}_2$ was present in $\tilde{\bm{d}_2}$, and how much information was obtained from $\hat{\bm{x}}_1$, the results of TCS-MRI and DELTA-MRI were compared against a reconstruction
obtained from $\tilde{\bm{d}_2}$ alone by applying an inverse DFT 
to the zero 
filled data $\tilde{\bm{d}_2}$. This method is denoted Z-IDFT. 

\textbf{Performance criterion:}
As a quantitative performance criterion, the normalized reconstruction error of the reconstructed image 
$\hat{\bm{r}}_2$ was calculated as: 
\begin{equation} \label{eqn:norm reconstruction error}
\epsilon(\hat{\bm r}_2) = \|\hat{\bm r}_2-\bm {r}_2 \|_2 / \|\hat{\bm {r}}_1-\bm{r}_2\|_2
    \quad ,
\end{equation}
where $\hat{\bm{r}}_2 := W(\hat{\bm r}_1, \hat{\bm \theta}, \hat{\bm t}, \hat{\bm v})$ for DELTA-MRI.

\section{Results \& Discussion}
\label{sec:results_discussion}
\begin{figure}[ht]
    \centering
    \begin{subfigure}[t]{2.8cm}
    \includegraphics[width=2.8cm]{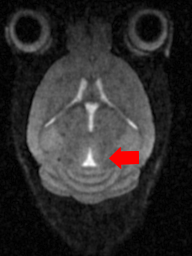}
    \caption{Z-IDFT}
    \label{fig:Direct_5proc_norm}
    \end{subfigure}
    \begin{subfigure}[t]{2.8cm}
    \includegraphics[width=2.8cm]{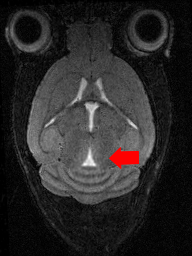}
    \caption{TCS-MRI}
    \label{fig:TCS-MRI_5proc_norm}
    \end{subfigure}
    \begin{subfigure}[t]{2.8cm}
    \includegraphics[width=2.8cm]{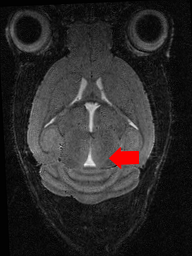}
    \caption{DELTA-MRI}
    \label{fig:Delta_MRI_5proc_norm}
    \end{subfigure}
    \newline
    \begin{subfigure}[t]{2.8cm}
    \includegraphics[width=2.8cm]{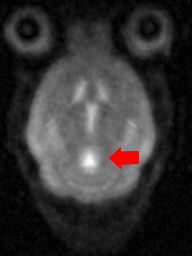}
    \caption{Z-IDFT}
    \label{fig:Direct_1proc_norm}
    \end{subfigure}
    \begin{subfigure}[t]{2.8cm}
    \includegraphics[width=2.8cm]{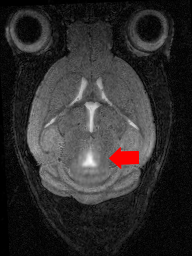}
    \caption{TCS-MRI}
    \label{fig:TCS_MRI_1porc_norm}
    \end{subfigure}
    \begin{subfigure}[t]{2.8cm}
    \includegraphics[width=2.8cm]{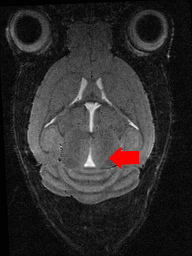}
    \caption{DELTA-MRI}
    \label{fig:Delta_MRI_1proc_norm}
    \end{subfigure}
    \caption{Reconstructions $\hat{\bm{r}}_2$ from 5\% (top) and 1\% (bottom) sub-sampled data.}
    \label{fig:rec_normal}
\end{figure}

\begin{figure}[!htb]
    \centering
    \includegraphics[width=8cm]{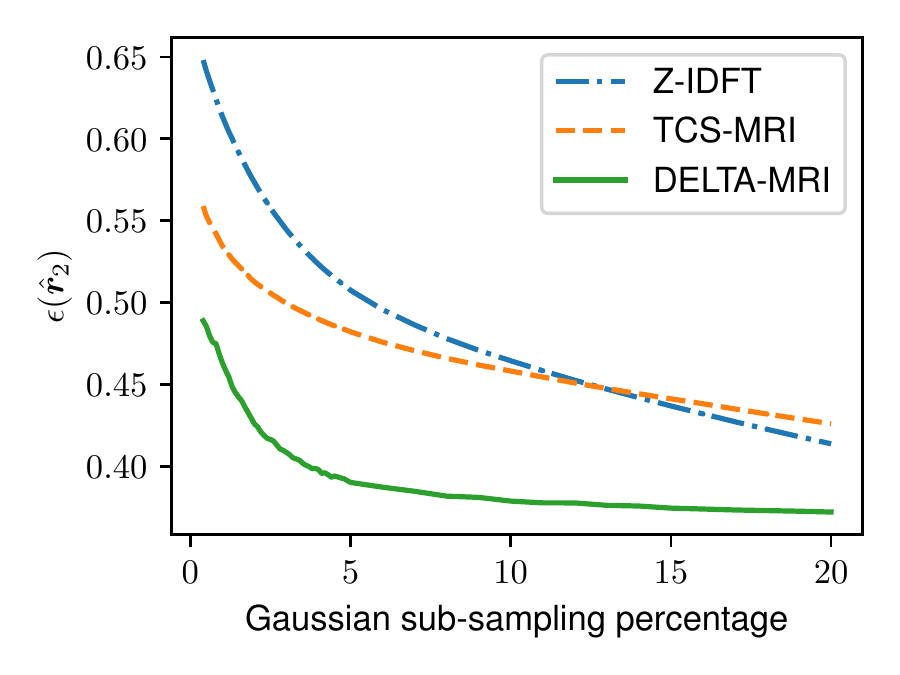}
    \caption{
    Normalized reconstruction error of Z-IDFT, TCS-MRI and DELTA-MRI as a function of the sub-sampling percentage of a Gaussian sub-sampling scheme.
    }
    \label{fig:quantitative comparison} 
\end{figure}

Figure \ref{fig:rec_normal}
shows reconstruction results obtained from Gaussian sub-sampled k-space data 
with sub-sampling percentages equal to 5\% and 1\%.
The left column shows 
(a cross-section of) the Z-IDFT reconstruction of $\bm{r}_2$.
The middle and right column show the reconstructions of $\bm{r}_2$ obtained by TCS-MRI and
DELTA-MRI, respectively. 
Figure \ref{fig:quantitative comparison} shows the normalized reconstruction error (\ref{eqn:norm reconstruction error}) as a function of the sub-sampling percentage for Z-IDFT, TCS-MRI and DELTA-MRI. 

The results show that for sub-sampling percentages up to 20\%, DELTA-MRI outperforms Z-IDFT and TCS-MRI
. Furthermore, \reffig{fig:quantitative comparison} shows that for a sub-sampling percentage below around 13\%, TCS-MRI outperforms Z-IDF, whereas  for higher sub-sampling percentages Z-IDF outperforms TCS-MRI. This
can be explained as follows. In our simulation study, the
weights of the regularization terms and data-misfit term that together constitute the cost function of \refeq{eqn:TCS-MRI} were kept 
proportional to each other 
and were tuned to be optimal for a sub-sampling percentage of 1\%. For increasing sub-sampling percentages, these regularization settings
lead to over-regularization, causing TCS-MRI to be outperformed by Z-IDF beyond a certain turning point. 
Alternatively, the regularization settings of  TCS-MRI and Delta-MRI could be optimized for each sub-sampling percentage individually, resulting in reduced reconstruction errors for all sub-sampling percentages excluding 1\%. However, this approach would not be realistic, as in practice ground truth images are not available and regularization settings are typically sub-optimal.

Additionally, some remarks can be made about the comparison between DELTA-MRI and TCS-MRI. First, unlike DELTA-MRI, TCS-MRI assumes the phases of $\bm{x}_1$ and $\bm{x}_2$ to be similar. This assumption is reflected by the first regularization term of \refeq{eqn:TCS-MRI}, which enforces similarity of $\bm{x}_1$ and $\bm{x}_2$ in the complex domain. In the simulation experiments, the ground truth phases $\bm{\phi}_1$ and $\bm{\phi}_2$ were chosen equal to meet this assumption, providing optimal conditions for  TCS-MRI.
In practice, however, the assumption of similar phase images may be
violated, which may negatively affect the performance of TCS-MRI. Second, as mentioned above, unlike DELTA-MRI, TCS-MRI does not include an alignment procedure, assuming that the rigid transformation aligning the images $\bm{x}_1$ and $\bm{x}_2$ is known beforehand. To allow a fair comparison that accounts for this difference, the rigid transformation estimated by DELTA-MRI was used to align the data before applying TCS-MRI. TCS-MRI could have been combined with alternative alignment procedures
, but this is not expected to change the main conclusions of the comparison.

The current work has some limitations that will be addressed in future work. First,  
while the current version of DELTA-MRI was designed for single-coil data, the method can be extended to account for multi-coil data, taking coil sensitivities into account. 
Furthermore, to facilitate experiments with prospectively under-sampled longitudinal data, the Gaussian distributed sampling schemes used to evaluate DELTA-MRI in this work can be replaced by sampling schemes that are practically more feasible. 

Finally, it is worthwhile mentioning that the problem of determining a DVF that deforms a source image to a target image is underdetermined. In other words, many DVFs may transform the source image to the same target image. This is known as the aperture problem \cite{Marr82,Horn93}. For this reason, in simulation experiments, the quality of a DVF estimate is not appropriately measured in terms of its deviation from the ground truth DVF, but in terms of the difference between the ground truth target image and the image that is obtained by geometrically deforming the source image according to the estimated DVF. This motivates the use of the reconstruction error \refeq{eqn:norm reconstruction error} as a performance criterion in this work.

\section{Conclusion}
\label{sec:conclusion}
We proposed DELTA-MRI, a novel framework that estimates longitudinal image changes directly from a reference image and strongly sub-sampled k-space data of a follow-up image.
The DVF and motion parameters that describe the relation between the reference image and the follow-up image are estimated in one-go. 
For sub-sampling percentages up to 20\%, DELTA-MRI was shown to outperform a state-of-the-art longitudinal CS based method in terms of reconstruction error.
\section{Compliance with ethical standards}
\label{sec:ethics}
This study was performed in line with the principles of the Declaration of Helsinki. All applicable institutional and/or national guidelines for the care and use of animals were followed. Procedures were performed in accordance with the European Directive 2010/63/EU on the protection of animals used for scientific purposes. The protocols were approved by the Committee on Animal Care and Use at the University of Antwerp, Belgium (permit number: 2016-10) and all efforts were made to minimize animal suffering. Experiments have been reported in compliance with the ARRIVE guidelines.
\section{Acknowledgments}
\label{sec:acknowledgments}
The authors thank J. van Rijswijk and J. Van Audekerke (Bio-Imaging Lab, Univ. Antwerp, Belgium) for the acquisition of the experimental MR data and valuable discussions. Jens Renders is an SB PhD fellow at the Research Foundation - Flanders (FWO), grant no. 1SA2920N. This work is partially supported by FWO project S007219N.

\begingroup
\raggedright

\bibliographystyle{IEEEbib-abbrev}
{\bibliography{refs}}
\endgroup

\end{document}